\documentclass[sigconf]{acmart}

\usepackage{booktabs} 
\usepackage{graphicx}
\usepackage{xcolor}
\usepackage{gensymb}
\usepackage{breakurl}
\usepackage{booktabs}
\usepackage{graphicx}
\usepackage[absolute,overlay]{textpos}

\renewcommand\footnotetextcopyrightpermission[1]{}

\copyrightyear{2022} 
\acmYear{2022} 
\setcopyright{acmlicensed}\acmConference[ASE '22]{37th IEEE/ACM International Conference on Automated Software Engineering}{October 10--14, 2022}{Rochester, MI, USA}
\acmBooktitle{37th IEEE/ACM International Conference on Automated Software Engineering (ASE '22), October 10--14, 2022, Rochester, MI, USA}
\acmPrice{15.00}
\acmDOI{10.1145/3551349.3561170}
\acmISBN{978-1-4503-9475-8/22/10}

\begin{document}
\sloppy

\begin{textblock*}{7cm}(1.5cm,26.2cm) 
	DOI: \url{https://doi.org/10.1145/3551349.3561170}
\end{textblock*}

\title{ESAVE: Estimating Server and Virtual Machine Energy}

\author{Priyavanshi Pathania$^\dagger$, Rohit Mehra$^\dagger$, Vibhu Saujanya Sharma$^\dagger$, Vikrant Kaulgud$^\dagger$, Sanjay Podder$^\ddagger$, Adam P. Burden*} 
\affiliation{ 
	\institution{$^\dagger$Accenture Labs, India \hspace{0.4em}
		$^\ddagger$Accenture, India \hspace{0.4em}
		*Accenture, USA}
	\country{}
}
\email{{{priyavanshi.pathania, rohit.a.mehra, vibhu.sharma, vikrant.kaulgud, sanjay.podder, adam.p.burden}@accenture.com}}

\begin{abstract}

Sustainable software engineering has received a lot of attention in recent times, as we witness an ever-growing slice of energy use, for example, at data centers, as software systems utilize the underlying infrastructure. Characterizing servers for their energy use accurately without being intrusive, is therefore important to make sustainable software deployment choices. In this paper, we introduce \textit{ESAVE} which is a machine learning-based approach that leverages a small set of hardware attributes to characterize a server or virtual machine's energy usage across different levels of utilization. This is based upon an extensive exploration of multiple ML approaches, with a focus on a minimal set of required attributes, while showcasing good accuracy. Early validations show that \textit{ESAVE} has only around 12\% average prediction error, despite being non-intrusive.

\end{abstract}

\keywords{Sustainable Software Engineering, Green IT, Energy Estimation, Carbon Emissions, Machine Learning, Energy Modeling}

\maketitle

\vspace{-0.3em}
\section{Introduction}\label{introduction}

While technology has been a great enabler for sustainability in various fields, it has a carbon footprint of its own. Multiple studies have estimated that the internet and communications technology industry (ICT) currently accounts for 2-7\% of the global greenhouse gas emissions (GHG), and is expected to increase to 14\% by 2040 \cite{unepccc}. A major share of these emissions can be attributed to the design, development, deployment, and distribution of software systems and the corresponding hardware infrastructure required to support these activities \cite{KERN201553}. Moreover, data centers (DC) required to run these software systems, currently account for 1-2\% of the global electricity demand (or about 0.3\% of the global GHG emissions) \cite{iea, nature}. If left unaddressed, these rapidly growing emissions can have a disastrous impact on the sustainability of our environment.

One of the primary reasons behind these high carbon emissions is the non-optimization of software systems from a sustainability perspective (specifically green/energy perspective) \cite{10.1145/3154384}. While there are multiple reasons like lack of awareness, intrusive approaches, hidden impact, etc. attributing to this non-optimization \cite{ICSE-SEIP}, the dearth of tools and techniques to quantify the environmental impact of a software system w.r.t the energy consumed by the underlying hardware, is one of the foremost \cite{10.1145/3154384, 7169668}. This acts as a major challenge for an environmentally-inclined practitioner (developer, project manager, etc.) who wants to adopt energy-efficient practices in her software, but does not have the required tooling support to measure and quantify the realized benefits. For example, quantifying the energy benefits of replacing an energy-hungry design pattern with an energy-efficient design pattern \cite{7203028}.

A majority of the industrial software systems are deployed on DC, including both on-premise and cloud DC. Also, in most cases, even the applications deployed on end-user devices (smartphone, laptop, desktop, etc.) usually have a backend component deployed on DC. Due to this sheer volume of compute happening on these DC, they generate almost double the emissions (45\% of ICT in 2020), in comparison to the end-user devices (24\% of ICT in 2020) \cite{BELKHIR2018448}. Hence, in order to have a bigger impact on the overall sustainability of our environment, optimizing software systems that reside on these DC, is the primary focus of this research.

In the recent past, there have been some tools and approaches to estimate the energy consumption of the underlying hardware (as a proxy for the energy consumption of the software system running on top), but they are either highly approximate, intrusive in nature, require access to a copious amount of static/run-time telemetry, etc. Although some of these approaches, along with their inherent limitations, might still serve the purpose of academic research, they are impractical from an industry usage standpoint. For example, the Cloud Carbon Footprint tool needs access to the cloud service provider's billing report, in order to estimate the energy consumption and carbon emissions of the utilized cloud resources \cite{ccf}. Granting access to such confidential information might not be possible for an organization due to adherence to multiple security, privacy, and compliance-related protocols. The problem further amplifies for a software services organization where project artifacts are predominantly owned by the clients. Keeping in mind these limitations, we believe that an \textit{ideal} energy estimation approach should (i) be non-intrusive (ii) leverage easily accessible telemetry (iii) support diverse architectures and manufacturers (iv) avoid constant retraining (v) estimate the energy of the entire hardware and not just components, while demonstrating (vi) high accuracy.

In this paper, we introduce \textit{ESAVE} (Estimating Server And Virtual machine Energy), a novel approach that enables energy estimation for software systems running on top of bare-metal servers or virtual machines (VM), for both on-premise and cloud DC. The approach is non-intrusive and leverages a user-friendly set of hardware configuration attributes and basic runtime telemetry to estimate the energy consumption using a custom-trained machine learning model, which has been trained on an industry-standard benchmark called \textit{SPECpower\_ssj2008}. Despite supporting non-intrusive estimation, along with other characteristics of an \textit{ideal} energy estimation approach, \textit{ESAVE} has been validated to exhibit less than 12\% average prediction error (across both cross and external validation).

\section{Formulating ESAVE}\label{esave}

For building a non-intrusive energy estimation approach, we leveraged the \textit{SPECpower\_ssj2008} dataset for training our ML models \cite{10.1007/978-3-540-69814-2_17, 10.1145/3185768.3185775}. It is the first industry-standard benchmark that measures the power of server class computers at graduated load levels of 0\% (idle) to 100\%, in 10\% increments. At the time of this research, the dataset had 755 observations corresponding to servers produced by noted original equipment manufacturers like IBM, HP, Dell, etc. and spread across 90 features describing the server configuration and its power characteristics. Please note that the dataset uses the terminology as load levels instead of CPU utilization levels, but for this paper, we have used them interchangeably \cite{5762739, 6665344}.

Post initial data cleanup, feature engineering, and extensive exploratory data analysis, we selected 15 features from the dataset to train our model on. These features describe the server characteristics pertaining to its CPU, memory, storage, and server nodes. Selection of these features was based upon a careful balance between statistical considerations (for e.g, the correlation between these features and server power characteristics) and ease-of-access considerations (for e.g, while power supply rating exhibits a decent correlation with server power characteristics, it is difficult to access, especially on the cloud, and hence not selected). Moreover, based upon our internal brainstorming sessions with key stakeholders in our organization, we devised another set of 8 reduced features, by giving higher importance to ease-of-access considerations, and selecting the features that describe the server characteristics pertaining only to its CPU. Table \ref{tab:table_cross_validation_results} lists both sets of selected features (ref. \textit{Models - All Features}) and reduced features (ref. \textit{Models - Reduced Features}). Finally, we selected 11 dependent variables (from power at 0\% (idle) CPU util. to power at 100\% CPU util., in increments of 10\%) to then train a set of 11 individual ML models for estimating each one of those dependent variables.

Since our cleaned-up dataset comprised mainly of continuous and dichotomous features, we leveraged regression-based algorithms to train our estimation models. Overall, we tested with 13 different regression algorithms (for example, linear regression, support vector regression, neural network regression, etc.).

To evaluate the goodness of fit for the trained models, we leveraged the Mean Absolute Percentage Error (MAPE) metric \cite{DEMYTTENAERE201638}. Out of the 13 regression algorithms that we tested, \textit{XGBoost} (an optimized gradient boosting algorithm) demonstrated the least MAPE, consistently across all 11 trained models \cite{10.1145/2939672.2939785}. Table \ref{tab:table_cross_validation_results} shows a summary of the 10-fold cross-validation results for the \textit{XGBoost} models. The average MAPE across all 11 models is reported to be 9.78\% for \textit{Models - All Features} and 12.27\% for \textit{Models - Reduced Features}.

The results indicate that the trained models can be used to estimate the power consumption curve of bare-metal servers by having access to the aforementioned set of hardware features, while demonstrating high accuracy. Moreover, the estimated power curve can thereafter be leveraged to estimate the energy consumption of a software system running on top of such a bare-metal server, by getting access to the average/temporal CPU utilization values recorded during a specific software run (using existing performance monitoring tools or simple batch/shell commands).

While these trained models currently only work for bare-metal servers, we are working on extending this approach to support VMs as well, by leveraging the concept of \textit{host energy apportioning}. This refers to apportioning of the underlying host's energy consumption to the respective VM, depending upon the portion of the host resources dedicated to the VM \cite{protocol2017ict}. Further explorations in this direction represent a major portion of our planned future work.

\begin{table}[]
\ttfamily
\footnotesize
\centering
\caption{10-fold cross-validation results for the \textit{XGBoost} models, across all 11 trained models. (P@CPU = Power at specific CPU Utilization, SE = Standard Error)}
\label{tab:table_cross_validation_results}
\vspace{-0.8em}
\resizebox{\columnwidth}{!}{%
\begin{tabular}{@{}l c c c c @{}}
\toprule
\textbf{}          & \multicolumn{2}{c}{\textbf{Models - All Features}}  & \multicolumn{2}{c}{\textbf{Models - Reduced Features}}    \\       
\textbf{}          & \multicolumn{2}{c}{\scriptsize \#Chips, \#CoresPerChip, \#ThreadsPerCore, CPUFrequency}  & \multicolumn{2}{c}{\scriptsize \#Chips, \#CoresPerChip, \#ThreadsPerCore, CPUFrequency}    \\ \addlinespace[-0.1ex]             
\textbf{}          & \multicolumn{2}{c}{\scriptsize L1CacheIValue, L1CacheDValue, L2Cache, L3Cache}           & \multicolumn{2}{c}{\scriptsize L1CacheIValue, L1CacheDValue, L2Cache, L3Cache}             \\ \addlinespace[-0.1ex]
\textbf{}          & \multicolumn{2}{c}{\scriptsize Memory, isSSD, \#StorageDrives, StorageCapacity}          & \multicolumn{2}{c}{\textbf{}}                                                                 \\ \addlinespace[-0.1ex]
\textbf{}          & \multicolumn{2}{c}{\scriptsize isSingleNode, isMultiNode, \#Nodes}                       & \multicolumn{2}{c}{\textbf{}}                                                                 \\ \cmidrule{1-1} \cmidrule(lr){2-3} \cmidrule(lr){4-5}
\textbf{P@CPU}     & \textbf{\,\,\,\,\,\,\,\,\, MAPE (\%) \,\,\,\,\,\,\,\,\,} & \textbf{SE (\%)}  & \textbf{\,\,\,\,\,\,\,\,\, MAPE (\%) \,\,\,\,\,\,\,\,\,} & \textbf{SE (\%)}                                                                                            \\ \cmidrule{1-1} \cmidrule(lr){2-3} \cmidrule(lr){4-5}
\textbf{P@idle}               & 16.35  & 1.15   & 20.69 & 3.33 \\
\textbf{P@10}                 & 10.50  & 0.74   & 13.35 & 2.11 \\
\textbf{P@20}                 & 10.31  & 1.05   & 12.55 & 2.09 \\
\textbf{P@30}                 & 9.39   & 0.61   & 11.89 & 2.06 \\
\textbf{P@40}                 & 9.13   & 0.53   & 11.78 & 2.02 \\
\textbf{P@50}                 & 8.96   & 0.55   & 11.64 & 1.89 \\
\textbf{P@60}                 & 8.74   & 0.49   & 11.37 & 1.81 \\
\textbf{P@70}                 & 8.68   & 0.52   & 10.73 & 1.67 \\
\textbf{P@80}                 & 8.39   & 0.49   & 10.68 & 1.74 \\
\textbf{P@90}                 & 8.39   & 0.48   & 10.48 & 1.67 \\
\textbf{P@100}                & 8.71   & 0.56   & 9.90  & 1.5  \\ \bottomrule
\end{tabular}%
}
\vspace{-1.5em}
\end{table}

\section{Early Validation}

To validate the effectiveness/accuracy of our approach, we conducted a set of early experiments that compared 
our energy estimation results to that of \textit{turbostat} \cite{turbostat}. Since \textit{turbostat} only works on bare-metal instances (VMs are not supported), we selected the Amazon AWS instance \textit{g4dn.metal} for our experiments. To exercise the instance, we trained three neural network-based ML models as workload. During the training process, the instance's energy was monitored using \textit{turbostat} and per second CPU utilization was logged using \textit{psutil} \cite{psutil}. Finally, an offline energy estimation was performed using \textit{ESAVE} by leveraging the hardware configuration of the bare-metal server and previously logged CPU utilization values (using the aforementioned \textit{Models - Reduced Features}). Table \ref{tab:table_validation_results} highlights the comparison between both the estimations. Despite \textit{ESAVE} being non-intrusive, the average deviation between both the estimations was recorded to be 7.89\%, which is very promising.

\begin{table}[]
\ttfamily
\centering
\caption{Early results of our validation experiments on comparing ESAVE and Turbostat.}
\label{tab:table_validation_results}
\vspace{-1.0em}
\resizebox{\columnwidth}{!}{%
\begin{tabular}{@{}l c c c c c @{}}
\toprule
\textbf{}                      & \textbf{}                                          & \multicolumn{1}{c}{\textbf{Turbostat}}           & \multicolumn{2}{c}{\textbf{ESAVE}}                                                                         & \textbf{}                  \\ \cmidrule(lr){3-3} \cmidrule(lr){4-5}
\textbf{Training Dataset}      & \textbf{Run Time (sec)}                            &  \textbf{Avg. Energy (J)}                        & \textbf{Avg. CPU Util. (\%)}                       & \textbf{Avg. Energy (J)}                              & \textbf{Diff. (\%)}        \\ \midrule

Kaggle Cats vs Dogs            &  599                                               &  127,587                                          & 63.82                                              & 113,810                                              & \textbf{11.41}              \\ 
Kaggle Natural Scenes          &  470                                               &  61,100                                           & 30.26                                              & 63,450                                               & \textbf{03.78}              \\ 
MNIST                          &  149                                               &  22,052                                           & 33.70                                              & 20,264                                               & \textbf{08.45}             \\ \bottomrule

\end{tabular}%
}
\vspace{-1.5em}
\end{table}

\section{Conclusion and Future Directions}\label{conclusion and future directions}

In this paper, we introduced \textit{ESAVE}, a novel approach for estimating the energy consumed by a software system, running on a server, by employing custom-trained ML models. Initial results indicate that our approach performs very well in terms of both cross and external validation. In the near future, we intend to extend \textit{ESAVE} to VMs, as well as conduct in-depth experiments to extensively validate its applicability/accuracy. \textit{ESAVE} is currently being piloted across multiple software sustainability use cases at our organization, with very encouraging initial feedback from key stakeholders.

\section*{Acknowledgement}

The authors would like to thank Samarth Sikand and Raghotham M. Rao for their invaluable support throughout the \textit{ESAVE} journey.

\bibliographystyle{ACM-Reference-Format}
\bibliography{Bibliography}


\begin{thebibliography}{19}


\ifx \showCODEN    \undefined \def \showCODEN     #1{\unskip}     \fi
\ifx \showDOI      \undefined \def \showDOI       #1{#1}\fi
\ifx \showISBNx    \undefined \def \showISBNx     #1{\unskip}     \fi
\ifx \showISBNxiii \undefined \def \showISBNxiii  #1{\unskip}     \fi
\ifx \showISSN     \undefined \def \showISSN      #1{\unskip}     \fi
\ifx \showLCCN     \undefined \def \showLCCN      #1{\unskip}     \fi
\ifx \shownote     \undefined \def \shownote      #1{#1}          \fi
\ifx \showarticletitle \undefined \def \showarticletitle #1{#1}   \fi
\ifx \showURL      \undefined \def \showURL       {\relax}        \fi
\providecommand\bibfield[2]{#2}
\providecommand\bibinfo[2]{#2}
\providecommand\natexlab[1]{#1}
\providecommand\showeprint[2][]{arXiv:#2}

\bibitem[Agency(2022)]%
        {iea}
\bibfield{author}{\bibinfo{person}{International~Energy Agency}.}
  \bibinfo{year}{Accessed - 20/07/2022}\natexlab{}.
\newblock \bibinfo{title}{Data Centres and Data Transmission Networks}.
\newblock
  \bibinfo{howpublished}{\url{https://www.iea.org/reports/data-centres-and-data-transmission-networks}}.
\newblock


\bibitem[Belkhir and Elmeligi(2018)]%
        {BELKHIR2018448}
\bibfield{author}{\bibinfo{person}{Lotfi Belkhir} {and} \bibinfo{person}{Ahmed
  Elmeligi}.} \bibinfo{year}{2018}\natexlab{}.
\newblock \showarticletitle{Assessing ICT global emissions footprint: Trends to
  2040 and recommendations}.
\newblock \bibinfo{journal}{\emph{Journal of Cleaner Production}}
  \bibinfo{volume}{177} (\bibinfo{year}{2018}), \bibinfo{pages}{448--463}.
\newblock
\showISSN{0959-6526}
\urldef\tempurl%
\url{https://doi.org/10.1016/j.jclepro.2017.12.239}
\showDOI{\tempurl}


\bibitem[Centre(2022)]%
        {unepccc}
\bibfield{author}{\bibinfo{person}{United Nations Environment Programme:
  Copenhagen~Climate Centre}.} \bibinfo{year}{Accessed -
  20/07/2022}\natexlab{}.
\newblock \bibinfo{title}{Greenhouse gas emissions in the ICT sector: trends
  and methodologies}.
\newblock
  \bibinfo{howpublished}{\url{https://c2e2.unepccc.org/wp-content/uploads/sites/3/2020/03/greenhouse-gas-emissions-in-the-ict-sector.pdf}}.
\newblock


\bibitem[Chen and Guestrin(2016)]%
        {10.1145/2939672.2939785}
\bibfield{author}{\bibinfo{person}{Tianqi Chen} {and} \bibinfo{person}{Carlos
  Guestrin}.} \bibinfo{year}{2016}\natexlab{}.
\newblock \showarticletitle{XGBoost: A Scalable Tree Boosting System}. In
  \bibinfo{booktitle}{\emph{Proceedings of the 22nd ACM SIGKDD International
  Conference on Knowledge Discovery and Data Mining}} (San Francisco,
  California, USA) \emph{(\bibinfo{series}{KDD '16})}.
  \bibinfo{publisher}{Association for Computing Machinery},
  \bibinfo{address}{New York, NY, USA}, \bibinfo{pages}{785–794}.
\newblock
\showISBNx{9781450342322}
\urldef\tempurl%
\url{https://doi.org/10.1145/2939672.2939785}
\showDOI{\tempurl}


\bibitem[cloudcarbonfootprint.org(2022)]%
        {ccf}
\bibfield{author}{\bibinfo{person}{cloudcarbonfootprint.org}.}
  \bibinfo{year}{Accessed - 20/07/2022}\natexlab{}.
\newblock \bibinfo{title}{Cloud Carbon Footrpint}.
\newblock \bibinfo{howpublished}{\url{https://www.cloudcarbonfootprint.org}}.
\newblock


\bibitem[{de Myttenaere} et~al\mbox{.}(2016)]%
        {DEMYTTENAERE201638}
\bibfield{author}{\bibinfo{person}{Arnaud {de Myttenaere}},
  \bibinfo{person}{Boris Golden}, \bibinfo{person}{Bénédicte {Le Grand}},
  {and} \bibinfo{person}{Fabrice Rossi}.} \bibinfo{year}{2016}\natexlab{}.
\newblock \showarticletitle{Mean Absolute Percentage Error for regression
  models}.
\newblock \bibinfo{journal}{\emph{Neurocomputing}}  \bibinfo{volume}{192}
  (\bibinfo{year}{2016}), \bibinfo{pages}{38--48}.
\newblock
\showISSN{0925-2312}
\urldef\tempurl%
\url{https://doi.org/10.1016/j.neucom.2015.12.114}
\showDOI{\tempurl}
\newblock
\shownote{Advances in artificial neural networks, machine learning and
  computational intelligence}.


\bibitem[Gray et~al\mbox{.}(2008)]%
        {10.1007/978-3-540-69814-2_17}
\bibfield{author}{\bibinfo{person}{Larry~D. Gray}, \bibinfo{person}{Anil
  Kumar}, {and} \bibinfo{person}{Harry~H. Li}.}
  \bibinfo{year}{2008}\natexlab{}.
\newblock \showarticletitle{Workload Characterization of the
  SPECpower{\_}ssj2008 Benchmark}. In \bibinfo{booktitle}{\emph{Performance
  Evaluation: Metrics, Models and Benchmarks}},
  \bibfield{editor}{\bibinfo{person}{Samuel Kounev}, \bibinfo{person}{Ian
  Gorton}, {and} \bibinfo{person}{Kai Sachs}} (Eds.).
  \bibinfo{publisher}{Springer Berlin Heidelberg}, \bibinfo{address}{Berlin,
  Heidelberg}, \bibinfo{pages}{262--282}.
\newblock
\showISBNx{978-3-540-69814-2}


\bibitem[Hsu and Poole(2011)]%
        {5762739}
\bibfield{author}{\bibinfo{person}{Chung-Hsing Hsu} {and}
  \bibinfo{person}{Stephen~W. Poole}.} \bibinfo{year}{2011}\natexlab{}.
\newblock \showarticletitle{Power signature analysis of the SPECpower\_ssj2008
  benchmark}. In \bibinfo{booktitle}{\emph{(IEEE ISPASS) IEEE International
  Symposium on Performance Analysis of Systems and Software}}.
  \bibinfo{pages}{227--236}.
\newblock
\urldef\tempurl%
\url{https://doi.org/10.1109/ISPASS.2011.5762739}
\showDOI{\tempurl}


\bibitem[Kern et~al\mbox{.}(2015)]%
        {KERN201553}
\bibfield{author}{\bibinfo{person}{Eva Kern}, \bibinfo{person}{Markus Dick},
  \bibinfo{person}{Stefan Naumann}, {and} \bibinfo{person}{Tim Hiller}.}
  \bibinfo{year}{2015}\natexlab{}.
\newblock \showarticletitle{Impacts of software and its engineering on the
  carbon footprint of ICT}.
\newblock \bibinfo{journal}{\emph{Environmental Impact Assessment Review}}
  \bibinfo{volume}{52} (\bibinfo{year}{2015}), \bibinfo{pages}{53--61}.
\newblock
\showISSN{0195-9255}
\urldef\tempurl%
\url{https://doi.org/10.1016/j.eiar.2014.07.003}
\showDOI{\tempurl}
\newblock
\shownote{Information technology and renewable energy - Modelling, simulation,
  decision support and environmental assessment}.


\bibitem[Lago(2015)]%
        {7169668}
\bibfield{author}{\bibinfo{person}{Patricia Lago}.}
  \bibinfo{year}{2015}\natexlab{}.
\newblock \showarticletitle{Challenges and Opportunities for Sustainable
  Software}. In \bibinfo{booktitle}{\emph{2015 IEEE/ACM 5th International
  Workshop on Product Line Approaches in Software Engineering}}.
  \bibinfo{pages}{1--2}.
\newblock
\urldef\tempurl%
\url{https://doi.org/10.1109/PLEASE.2015.8}
\showDOI{\tempurl}


\bibitem[linux.org(2022)]%
        {turbostat}
\bibfield{author}{\bibinfo{person}{linux.org}.} \bibinfo{year}{Accessed -
  20/07/2022}\natexlab{}.
\newblock \bibinfo{title}{TURBOSTAT}.
\newblock
  \bibinfo{howpublished}{\url{https://www.linux.org/docs/man8/turbostat.html}}.
\newblock


\bibitem[Mehra et~al\mbox{.}(2022)]%
        {ICSE-SEIP}
\bibfield{author}{\bibinfo{person}{Rohit Mehra},
  \bibinfo{person}{Vibhu~Saujanya Sharma}, \bibinfo{person}{Vikrant Kaulgud},
  \bibinfo{person}{Sanjay Podder}, {and} \bibinfo{person}{Adam~P. Burden}.}
  \bibinfo{year}{2022}\natexlab{}.
\newblock \showarticletitle{Towards a Green Quotient for Software Projects}. In
  \bibinfo{booktitle}{\emph{To appear in the Proceedings of the 40th
  International Conference on Software Engineering: Software Engineering in
  Practice}} (Pittsburgh, Pennsylvania, USA) \emph{(\bibinfo{series}{ICSE
  '22})}. \bibinfo{publisher}{ACM}.
\newblock
\urldef\tempurl%
\url{https://doi.org/10.48550/arXiv.2204.12998}
\showDOI{\tempurl}


\bibitem[nature.com(2022)]%
        {nature}
\bibfield{author}{\bibinfo{person}{nature.com}.} \bibinfo{year}{Accessed -
  20/07/2022}\natexlab{}.
\newblock \bibinfo{title}{How to stop data centres from gobbling up the
  world’s electricity}.
\newblock
  \bibinfo{howpublished}{\url{https://www.nature.com/articles/d41586-018-06610-y}}.
\newblock


\bibitem[Noureddine and Rajan(2015)]%
        {7203028}
\bibfield{author}{\bibinfo{person}{Adel Noureddine} {and}
  \bibinfo{person}{Ajitha Rajan}.} \bibinfo{year}{2015}\natexlab{}.
\newblock \showarticletitle{Optimising Energy Consumption of Design Patterns}.
  In \bibinfo{booktitle}{\emph{2015 IEEE/ACM 37th IEEE International Conference
  on Software Engineering}}, Vol.~\bibinfo{volume}{2}.
  \bibinfo{pages}{623--626}.
\newblock
\urldef\tempurl%
\url{https://doi.org/10.1109/ICSE.2015.208}
\showDOI{\tempurl}


\bibitem[Pinto and Castor(2017)]%
        {10.1145/3154384}
\bibfield{author}{\bibinfo{person}{Gustavo Pinto} {and}
  \bibinfo{person}{Fernando Castor}.} \bibinfo{year}{2017}\natexlab{}.
\newblock \showarticletitle{Energy Efficiency: A New Concern for Application
  Software Developers}.
\newblock \bibinfo{journal}{\emph{Commun. ACM}} \bibinfo{volume}{60},
  \bibinfo{number}{12} (\bibinfo{date}{nov} \bibinfo{year}{2017}),
  \bibinfo{pages}{68–75}.
\newblock
\showISSN{0001-0782}
\urldef\tempurl%
\url{https://doi.org/10.1145/3154384}
\showDOI{\tempurl}


\bibitem[Protocol(2017)]%
        {protocol2017ict}
\bibfield{author}{\bibinfo{person}{GHG Protocol}.}
  \bibinfo{year}{2017}\natexlab{}.
\newblock \showarticletitle{ICT Sector Guidance built on the GHG Protocol
  Product Life Cycle Accounting and Reporting Standard}.
\newblock \bibinfo{journal}{\emph{Global: GHG Protocol}}
  (\bibinfo{year}{2017}).
\newblock


\bibitem[pypi.org(2022)]%
        {psutil}
\bibfield{author}{\bibinfo{person}{pypi.org}.} \bibinfo{year}{Accessed -
  20/07/2022}\natexlab{}.
\newblock \bibinfo{title}{psutil}.
\newblock \bibinfo{howpublished}{\url{https://pypi.org/project/psutil/}}.
\newblock


\bibitem[von Kistowski et~al\mbox{.}(2018)]%
        {10.1145/3185768.3185775}
\bibfield{author}{\bibinfo{person}{J\'{o}akim von Kistowski},
  \bibinfo{person}{Klaus-Dieter Lange}, \bibinfo{person}{Jeremy~A. Arnold},
  \bibinfo{person}{Sanjay Sharma}, \bibinfo{person}{Johann Pais}, {and}
  \bibinfo{person}{Hansfried Block}.} \bibinfo{year}{2018}\natexlab{}.
\newblock \showarticletitle{Measuring and Benchmarking Power Consumption and
  Energy Efficiency}. In \bibinfo{booktitle}{\emph{Companion of the 2018
  ACM/SPEC International Conference on Performance Engineering}} (Berlin,
  Germany) \emph{(\bibinfo{series}{ICPE '18})}. \bibinfo{publisher}{Association
  for Computing Machinery}, \bibinfo{address}{New York, NY, USA},
  \bibinfo{pages}{57–65}.
\newblock
\showISBNx{9781450356299}
\urldef\tempurl%
\url{https://doi.org/10.1145/3185768.3185775}
\showDOI{\tempurl}


\bibitem[Zhang et~al\mbox{.}(2013)]%
        {6665344}
\bibfield{author}{\bibinfo{person}{Xiao Zhang}, \bibinfo{person}{Jianjun Lu},
  {and} \bibinfo{person}{Xiao Qin}.} \bibinfo{year}{2013}\natexlab{}.
\newblock \showarticletitle{BFEPM: Best Fit Energy Prediction Modeling Based on
  CPU Utilization}. In \bibinfo{booktitle}{\emph{2013 IEEE Eighth International
  Conference on Networking, Architecture and Storage}}.
  \bibinfo{pages}{41--49}.
\newblock
\urldef\tempurl%
\url{https://doi.org/10.1109/NAS.2013.12}
\showDOI{\tempurl}


\end{thebibliography}

\end{document}